# Using Data Science in High School Astronomy

James Newland[1]

[1]*Texas Advanced Computing Center, University of Texas at Austin, Texas, USA*

**Abstract.** Astronomy datasets can be challenging to use for high school astronomy classes. Data science education pedagogy can be leveraged to create astronomy activities in which students interrogate data, create visuals, and use statistical thinking to construct astronomy knowledge. This session describes how the NASA/IPAC Infrared Science Archive (IRSA) can provide a web-based interface for students to use basic data science techniques in astronomy to build data literacy while learning astronomical concepts. The activities shared will be available for anyone but were designed to be used in astro 101 classes in high school or early college.

## 1. Introduction

Modern astronomy has become a "big data" discipline (Lundgren & Trainor, 2023; Norman et al., 2019; Rebull, 2024; Taghizadeh-Popp et al., 2020). Bringing computing and data science pedagogy to current students is one way to support the development of the next generation of early-career scientists (Norman et al., 2019; Lundgren & Trainor, 2023). This trend is already evident in undergraduate physics programs, where computing is being integrated into the curriculum (Apple et al., 2021; Hutchins et al., 2020; Lee et al., 2020; Orban & Teeling-Smith, 2020; Weintrop et al., 2016; Weller et al., 2021). Data science and computing skills can also be a part of astronomy at the high school level. Work is already happening to incorporate modern data science (Bargagliotti et al., 2020; Israel-Fishelson et al., 2024) and computational thinking (Orban & Teeling-Smith, 2020; Weintrop et al., 2016) into high school science courses such as astronomy and physics (Norman et al., 2019; Newland, 2020; Rebull, 2024).

The NASA/IPAC Infrared Science Archive has become a sophisticated but accessible interface for publicly available datasets (Rebull, 2024). The development of the Firefly framework by NASA/IPAC (Wu et al., 2019) allows large data tables, rendered FITS images, and embedded plots to coexist in a single browser window. The IRSA tools present data using a table interface, which uses filtering techniques, including mathematical expressions. Data from the table can be immediately rendered as a plot, including visualization elements where variables can lead to color or size variations within a distribution. Visualizations can help the viewer interpret the meaning of variability within a distribution by mapping a quantity onto size or color (Schloss et al., 2019).

The activity presented here is meant to be incorporated into a high school astronomy or physics course. Rather than having students work through an entire curriculum related to data science, integrating data science pedagogy into data-driven courses, like astronomy, means that students can construct knowledge about astronomy by using computing in a specific context (Newland, 2020; Guzdial & Shreiner, 2022). The goal of the activity is for students to learn how stellar data from the Gaia spacecraft can be used to answer questions about various Milky Way star clusters.

The publication of the Gaia Data Release 2 paper describing observational Herzsprung-Russell diagrams (HRD) using the newly released data offered some excellent known targets (Gaia Collaboration, 2018) to use in developing the activity presented here. HRD creation using photometry from the Gaia pipeline often uses the calculated color index as an analog for temperature on the horizontal axis, and absolute magnitude on the vertical axis. The photometry and parallax values from the Gaia data pipeline can be used to determine the absolute magnitude. Both open and globular clusters were selected for the activity so that the stellar distributions were varied and the evolutionary tracks of the clusters would stand out. The plots produced as a part of the activity are color-magnitude diagrams (CMD), a subset of the HRD typically shown in astro 101.

## 2. Data Science Pedagogy in Astronomy

Data has become a significant part of life, and the world is becoming data-driven. Giving K-12 students opportunities to learn foundational data science concepts is more important now than ever (Israel-Fishelson et al., 2024). The Pre-K-12 Guidelines for Assessment and Instruction in Statistics Education II (GAISE II) framework describes a statistical problem-solving process (Bargagliotti et al., 2020) that can help guide the design of data science activities in K-12 classrooms.

Data science is best described in this context as using computing and statistics to tell a story using data. Data science pedagogy is part of the more extensive set of computational thinking skills science students should develop. The activity has been used in a high school "Astronomy 101" course over several iterations. These were typical astronomy students, and their math expertise was mixed at best. Most students completing this activity did not have extensive computing experience, including only introductory skills with spreadsheets.

### 2.1. Statistical Problem-Solving Practice

Students start by formulating investigative questions using statistical thinking. For the star cluster activity using the NASA/IPAC IRSA Gaia catalog, students want to know the relative ages of the clusters, the distances of the clusters, and a sense of the size of the cluster. The rigor level of the activity could be altered by having the students determine research questions on their own. The presented activity uses a scaffolded guided inquiry style where questions are suggested or given.

Next, students collect or consider existing data that helps them address the questions they asked. Students learn to query the catalog via the IRSA interface using appropriate nomenclature for the clusters based on the SIMBAD service (Wenger et al., 2000), which also serves as the source for proper angular sizes for the clusters to limit the number of returned sources from the catalog.

Then, students analyze data, including reducing the data into a functional form. The data from the catalogs to the IRSA tools uses the Firefly framework (Wu et al., 2019), producing interactive data tables with filtering capabilities. Students can use simple mathematical statements to limit the parallaxes displayed and processed in the browser window. Providing scaffolded feedback to students about appropriate filtering techniques helps to lower the cognitive load on students new to using data science tools on real datasets.

Finally, students interpret and communicate their results, sharing possible alternative explanations. Looking for alternative explanations for a given dataset is crucial in applying statistical thinking in the classroom (Bargagliotti et al., 2020; Israel-Fishelson et al., 2024). Students can learn to develop the needed critical view of the role of data when interpreting research findings. Using a critical lens for research is a data science skill that extends beyond the classroom into the wider world.

## 2.2. Task-Specific Computing for Data Science

Table 1 lists the clusters and the parameters students will use to find and filter the data. The angular sizes and parallax ranges were retrieved from the SIMBAD service (Wenger et al., 2000). The rigor level of the activity can be adjusted by having students determine these values themselves using SIMBAD. The activity was designed for introductory astronomy students to complete in a one-hour class session. Providing the parameters as shown is another way the activity can be scaffolded for students.

The activity's use of scaffolding means data science is task-specific, and thus, students are only exposed to the computing knowledge required to achieve the given goal. Using curricular tools to bring data science into a given discipline for a given task can help bolster (Guzdial & Shreiner, 2022) the addition of computational thinking to astronomy courses.

When teachers see a given technology as beneficial to their teaching practice, they are more likely to adopt it. Scaffolding in the activity is not only for the sake of time; task-specific computing used in a data science context can help lower the cognitive load on learners (Grover, 2022), especially those new to using computing to answer statistical questions in a science class.

Table 1. Cluster Parameters for HRD Creation

| Cluster Name | Angular Size (arcminutes) | Parallax Range (mas) | Cluster Type |
|---|---|---|---|
| Collinder 110 | 18 | > 0.30 and < 0.50 | Open |
| NGC 4755 | 10 | > 0.35 and < 0.50 | Open |
| Messier 13 | 20 | > 0.10 and < 0.20 | Globular |
| Melotte 71 | 9 | > 0.30 and < 0.50 | Open |

Lastly, the Gaia holdings as a part of the NASA/IPAC IRSA database collection could be considered Big Data. The IRSA Firefly (Wu et al., 2019) visualization tools are designed to group data into bins when data tables contain more than 25,000 rows. In order to produce the sort of CMD used in Astronomy 101, the tables need to be culled down using data filtering in the Firefly tables. Students will not produce valuable results without scaffolding and guidance limiting the data, likely impacting their knowledge construction.

## 3.  Creating Color-Magnitude Diagrams using the IRSA Gaia Database

### 3.1. Helpful Prior Knowledge

The activity centers around students using statistical problem-solving to characterize certain aspects of the star clusters being investigated. Some concepts would be helpful for students to experience before trying to handle the activity. For example, students must use right ascension, declination, and angular size in context. Understanding how astronomical images are produced would also probably bolster students' understanding of the representations supplied by the IRSA tools of infrared image data.

The general goal of the activity is for students to use statistical thinking and data science pedagogy to apply knowledge about how astronomers determine the ages of clusters and how cluster stellar populations are characterized. The activity is meant to apply students' knowledge of the differences between globular clusters and open clusters, how to use HRD and CMD to characterize stellar populations, how to use apparent and absolute magnitude, and how to use the distance modulus relationship algebraically. Experience interpreting HRD and CMD representations would be a helpful skill to practice before trying the activity in class.

Lastly, this activity might be helpful as part of a series of data-driven astronomy activities but probably not introduce the concept of working with tables of data. Data science educators acknowledge that learning to work with data in tabular form is essential to applying statistical thinking (Dana Center at UT Austin, 2024; Israel-Fishelson et al., 2024; Bargagliotti et al., 2020). Considering that IRSA Firefly tables are designed for scientific use rather than classroom use, students should likely have used tools like spreadsheets to manipulate and plot data before completing this activity.

When this activity was used with high school astronomy students, the classes had already done several activities using spreadsheets to manipulate and visualize data. The activity was used during the expected scope and sequence of high school astronomy as a part of a unit about stellar clusters in the Milky Way galaxy. Although the activity was designed to be completed in a single hour-long class session, this kind of activity often took two sessions for most students. Scaffolding for data science concepts was also provided frequently. The takeaway was that students learning how data can help us characterize clusters was more important than students learning data science pedagogy deeply.

### 3.2. Using IRSA Tools to Create Color-Magnitude Diagrams

Before creating a CMD with Gaia photometry data, students will use the NASA/IPAC IRSA Viewer interface (Wu et al., 2019) to search for targets. The IRSA Viewer interface allows users to enter a target name that is resolved using a connection to the SIMBAD astronomical database (Wenger et al., 2000). The nomenclature used to identify the targets given to the students was tested in the SIMBAD system before being used in the classroom for the first time. SIMBAD was also used to confirm parallax ranges for the clusters in the activity. Figure 1 shows the NASA/IPAC IRSA Viewer interface for one of the targets from the activity.

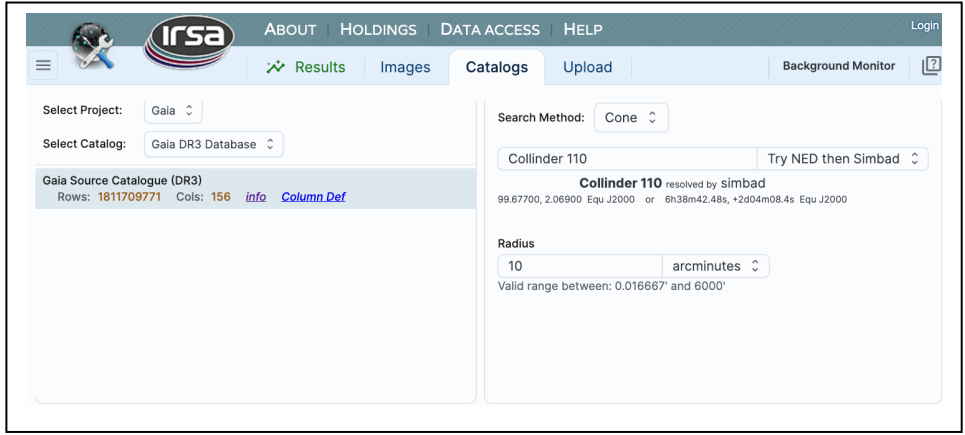

Figure 1 – NASA/IPAC IRSA Viewer Search Interface for Collinder 110

The sources returned from the Gaia catalog appear as a Firefly table (Wu et al., 2019). An image of the sources overplotted on the sky is also returned and displayed. Figure 2 shows the results for one of the targets from the activity. Imagery from the 2MASS survey is seen in this example. The IRSA Firefly JavaScript library allows large datasets to be displayed in a browser window with a separate data table, accompanying imagery with source locations overlayed, and a plot of the data using fields from the catalog queried. Large datasets of thousands of rows are possible when using the IRSA catalog offerings; therefore, a slower or older computer might take longer to display results.

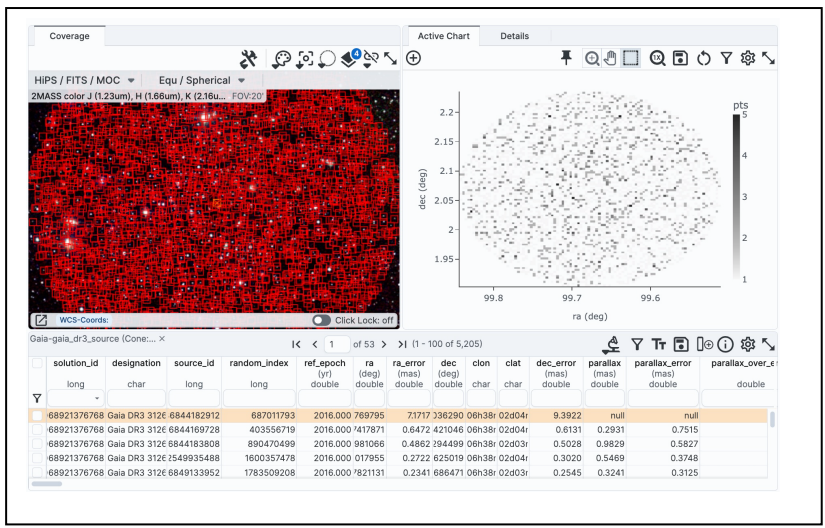

Figure 2 – NASA/IPAC IRSA Viewer Output for Collinder 110

The IRSA Firefly interface allows the user to make the plot pane the sole focus, which makes the production of a CMD easier. The plot parameters dialog allows the user to access any field from the catalog and some mathematical and logical functions. The Gaia data pipeline provides the color index using the magnitudes in blue, green, and red in a few combinations, making a good choice for the horizontal axis for a CMD (Gaia Collaboration, 2019). The absolute magnitude can be calculated and used for the vertical axis. This formula comes from the distance modulus relationship: **phot_g_mean_mag - (5 - 5*log10(1000/parallax))**. The expression uses the syntax that IRSA Firefly understands and the Gaia database's proper field names.

Providing students with the exact syntax and field names acts as scaffolding so that students can produce a CMD that they can interpret. The activity aims to have students perform sense-making tasks rather than teach them how to use the IRSA Viewer tools for professional astronomy. Field testing of the activity showed that focusing on the learning outcomes related to clusters led to more success than concerns over the lack of competence in using the NASA/IPAC IRSA interface for future work.

Figure 3 shows an example of a CMD with appropriate axes and plot labeling. The **bp_rp** color index field from the Gaia database was used for the horizontal axis. The colored points demonstrate one final potential application of data science pedagogy to separate stars within the distribution by parallax. Visualizations within a data distribution allow for an indirect "third axis," allowing students to perform deeper sense-making with the plots (Schloss et al., 2019).

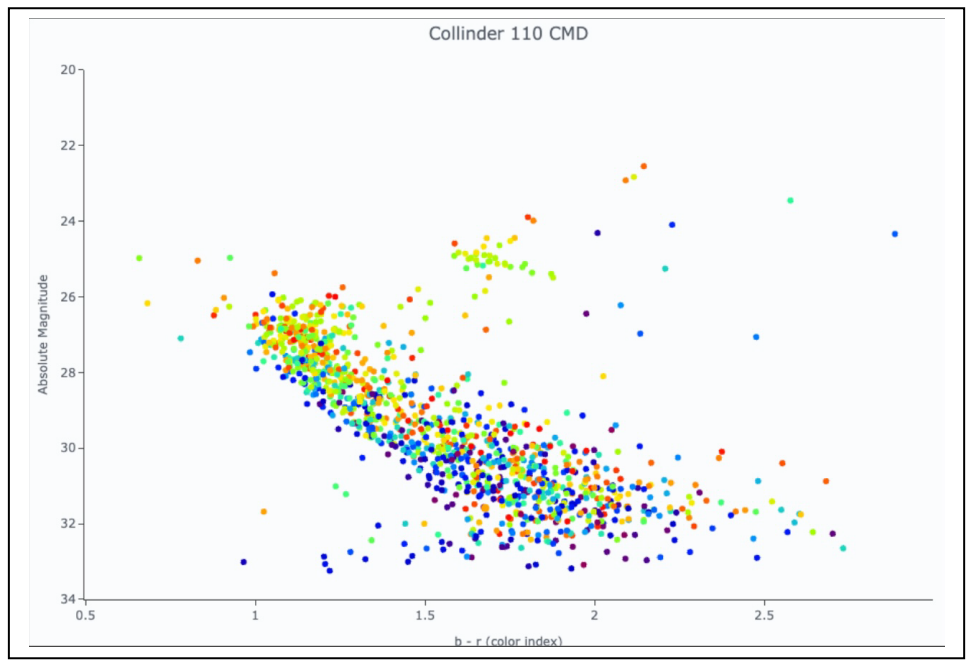

Figure 3 – Color Magnitude Diagram Produce using IRSA Viewer for Collinder 110

The activity shared here uses open-ended questions for an overall inquiry-driven approach, using CMDs to compare and contrast the clusters. Exploring cluster parameters such as relative age by inspecting the main sequence turn-off, distance by known parallax values, cluster size through analysis of the parallax range, etc., is all possible for an "Astronomy 101" class with a data-driven activity such as the one explored here. The questions are meant to have students explore the application of HRDs and knowledge about cluster evolution and stellar populations. Using authentic data and well-researched data science pedagogy allows students to apply past learning about clusters to knowledge construction through sense-making while using a professional astronomical tool with the proper guardrails in place.

The complete activity and other resources are available at the author's website (JimmyNewland.com). All curricular activities are free for classroom use under the provided Creative Commons license.

**Acknowledgements**. The author would like to acknowledge the help and support of Dr. Luisa Rebull from the NASA/IPAC Infrared Science Archive. This activity would not have been possible without her guidance and direction. In addition, the author would like to acknowledge the educators who collaborated as a part of the Big NITARP Alumni Project (BINAP) to create material to bring astronomical research into the classroom. This activity grew from collaborations with BINAP educators and was only possible with their help.